\documentclass[12pt,preprint]{aastex}

\input epsf

\shorttitle{Excitation and damping of $p$ modes of $\alpha$\,Cen~B}
\shortauthors{Chaplin et al.}

\begin{document}

\title{Excitation and damping of $p$-mode oscillations of
$\alpha$\,Cen~B}

\author{W.~J.~Chaplin} \affil{School of Physics and Astronomy,
University of Birmingham, Edgbaston, Birmingham B15 2TT, UK}
\email{w.j.chaplin@bham.ac.uk}

\and

\author{G.~Houdek} \affil{Institute of Astronomy, University of
Cambridge, Cambridge CB3 0HA, UK} \email{hg@ast.cam.ac.uk}

\and

\author{Y.~Elsworth} \affil{School of Physics and Astronomy,
University of Birmingham, Edgbaston, Birmingham B15 2TT, UK}

\and

\author{R.~New} \affil{Faculty of Arts, Computing, Engineering and
Sciences, Sheffield Hallam University, Sheffield S1 1WB, UK}

\and

\author{T. R. Bedding} \affil{School of Physics, University of Sydney,
Sydney, NSW 2006, Australia}

\and

\author{H. Kjeldsen} \affil{Department of Physics and Astronomy,
University of Aarhus, DK-8000 Aarhus C, Denmark}

\begin{abstract}

This paper presents an analysis of observational data on the p-mode
spectrum of the star $\alpha$\,Cen~B, and a comparison with
theoretical computations of the stochastic excitation and damping of
the modes. We find that at frequencies $\ga 4500\,\rm \mu Hz$, the
model damping rates appear to be too weak to explain the observed
shape of the power spectral density of $\alpha$\,Cen~B. The conclusion
rests on the assumption that most of the disagreement is due to
problems modelling the damping rates, not the excitation rates, of the
modes. This assumption is supported by a parallel analysis of BiSON
Sun-as-a-star data, for which it is possible to use analysis of very
long timeseries to place tight constraints on the assumption. The
BiSON analysis shows that there is a similar high-frequency
disagreement between theory and observation in the Sun.

We demonstrate that by using suitable comparisons of theory and
observation it is possible to make inference on the dependence of the
p-mode linewidths on frequency, without directly measuring those
linewidths, even though the $\alpha$\,Cen~B dataset is only a few
nights long. Use of independent measures from a previous study of the
$\alpha$\,Cen~B linewidths in two parts of its spectrum also allows us
to calibrate our linewidth estimates for the star. The resulting
calibrated linewidth curve looks similar to a frequency-scaled version
of its solar cousin, with the scaling factor equal to the ratio of the
respective acoustic cut-off frequencies of the two stars. The ratio of
the frequencies at which the onset of high-frequency problems is seen
in both stars is also given approximately by the same scaling factor.

\end{abstract}


\section{Introduction}
\label{sec:intro}

Stars like the Sun, which have sub-surface convection zones, display a
rich spectrum of acoustic (p-mode) oscillations. The oscillations are
stochastically excited and damped by the convection, and this gives
rise to an extremely rich spectrum of modes. Measurement of the
amplitudes and damping rates of the p modes therefore gives important
information for constraining theories of convection in stellar
interiors.

Chaplin et al. (2007) used model computations of the excitation and
damping of Sun-like oscillations to make predictions of the p-mode
spectra of a selection of stars on the lower main sequence.  The
computations revealed an extremely interesting feature in the
predicted appearance of the spectra of those model stars that had
effective temperatures cooler than about 5400\,K: the modelled power
spectral density of the modes showed two maxima, at different
frequencies. Chaplin et al. found a pronounced dip in mode power
between the maxima when the computations were made for young stars;
since the maxima are well separated in frequency, the predicted
spectra took on a ``double humped'' appearance. In older main-sequence
stars the dip was found to be much less pronounced, and instead the
spectra showed a broad plateau of power.

The K1~V main-sequence star $\alpha$\,Cen~B (HR\,5460) is a suitable
candidate to test the predictions, since its effective temperature
lies on the cool side of the 5400-K boundary given by the model
computations, and data are available on its p-mode spectrum from
observations made by Kjeldsen et al. (2005). As we shall demonstrate,
the model computations predict a broad plateau of power for
$\alpha$\,Cen~B; however, this broad plateau is not seen in the p-mode
data of Kjeldsen et al.  This paper reports on attempts to try and
understand this disagreement between theory and observation.

The observational data for our study are the aforementioned Doppler
velocity observations of $\alpha$\,Cen~B, made by Kjeldsen et al.  The
theoretical predictions we use are pulsation computations of the
stochastic excitation rates and the damping rates of the radial p
modes of the star. By making judicious comparisons of the theoretical
computations and the observational data we show it is actually
possible to make inference on the p-mode linewidths of
$\alpha$\,Cen~B, without directly measuring those linewidths, even
though the Kjeldsen et al. observations span only a few nights. The
conclusions drawn do rest on one important assumption: that most of
the disagreement that is observed between theoretical predictions and
observations of the radial-mode amplitudes of $\alpha$\,Cen~B resides
in errors in the model damping rates, and not errors in the model
excitation rates. We provide evidence in support of this
rationalization from a similar comparison of theoretical and
observational data of the p-mode spectrum of the ``Sun as a
star''. The Sun-as-a-star data come from observations made by the
ground-based Birmingham Solar-Oscillations Network (BiSON) (Chaplin et
al. 1996) in Doppler velocity.

The layout of our paper is as follows. We begin in
Section~\ref{sec:data} with a description of the excitation and
damping-rate computations. We then proceed in Section~\ref{sec:res} to
compare theoretical predictions and observations of the mode
amplitudes, for both $\alpha$\,Cen~B and the Sun. Then, in
Section~\ref{sec:wid}, we show how discrepancies between the
theoretical and observed amplitudes may be explained largely in terms
of errors in the theoretically computed damping rates. We also
demonstrate how inference may be made on the mode linewidths from use
of the observed amplitudes and the theoretical excitation and damping
computations. We finish in Section~\ref{sec:disc} with a brief
discussion of the main points of the paper.

 \section{Predictions from analytical model computations}
 \label{sec:data}

The stellar equilibrium and pulsation computations that we performed
are as described by Balmforth (1992a) and Houdek et. al (1999).  These
computations gave estimates of the powers and damping rates of the
radial p modes of $\alpha$\,Cen~B.

The model computations required four general input parameters to
specify the stellar model: the mass, $M$; radius, $R$; effective
temperature, $T_{\rm eff}$; and chemical composition. We took values
for $\alpha$\,Cen~B of: $M=0.934 \pm 0.007\,\rm M_{\odot}$; $R=0.863
\pm 0.005\,\rm R_{\odot}$; and $T_{\rm eff} = 5288 \pm 38\,\rm K$ (see
Yildiz 2007, and references therein). The composition was fixed at
$X=0.7$ and $Z=0.02$, as for the model computations performed in
Chaplin et al. (2005, 2007). While this composition differs from
estimates given in the literature for the $\alpha$\,Cen~B system (of
$Z \approx 0.025$; again, see Yildiz 2007), changes in composition at
this level have only a second-order impact on the model calculated
p-mode excitation and damping rates, and any changes that might be
relevant here are smaller than the observational uncertainties (e.g.,
see Fig.~17 of Houdek et al. 1999).

 \subsection{Pulsation computations}
 \label{sec:comp}  

Computation of the excitation rates and damping rates of the p modes
demands a description of how the pulsations interact with the
convection. This requires computation of the turbulent fluxes
associated with the convective heat and momentum transport. These
turbulent fluxes are obtained from a nonlocal, time-dependent
generalization of the mixing-length formulation of Gough (1977a, b),
with a mixing length calibrated to the Sun. In this generalization
there is a parameter, $\Phi$, which specifies the shape of the
convective eddies. Then there are two parameters, $a$ and $b$, which
control respectively the spatial coherence of the ensemble of eddies
contributing to the turbulent fluxes of heat and momentum and the
degree to which the turbulent fluxes are coupled to the local
stratification. These two parameters control the degree of
`non-locality' of convection; low values imply highly nonlocal
solutions, and in the limit $a,b\rightarrow\infty$ the system of
equations formally reduces to the local formulation (except near the
boundaries of the convection zone, where the local equations are
singular). Gough (1977a) has suggested theoretical estimates for their
values, but it is likely that the standard mixing-length assumption of
assigning a unique scale to turbulent eddies at any given location
causes too much smoothing; accordingly, somewhat larger values
probably yield more realistic results. In this paper we therefore
adopt two sets of values for the non-local parameters which provide
reasonable results for the Sun, and other Sun-like stars:
$a^2=b^2=600$; and $a^2=b^2=300$. We comment further on the impact of
this choice in Section~\ref{sec:cal} below.

 \subsubsection{Envelope and pulsation models}
 \label{sec:envpuls}

Both the envelope and pulsation computations assumed the
three-dimensional Eddington approximation to radiative transfer (Unno
\& Spiegel 1966).  The integration was carried out inwards, starting
at an optical depth of $\tau$=$10^{-4}$ and ending at a radius
fraction $r/R=0.2$.  The opacities were obtained from the OPAL tables
(Iglesias \& Rogers 1996), supplemented at low temperature by tables
from Kurucz (1991).  The equation of state included a detailed
treatment of the ionization of C, N, and O, and a treatment of the
first ionization of the next seven most abundant elements
(Christensen-Dalsgaard 1982), as well as `pressure ionization' by the
method of Eggleton, Faulkner \& Flannery (1973); electrons were
treated with relativistic Fermi-Dirac statistics.  Perfectly
reflective mechanical and thermal outer boundary conditions in the
pulsation computation were applied at the temperature minimum in the
manner of Baker \& Kippenhahn (1965). At the base of the model
envelope the conditions of adiabaticity and vanishing displacement
were imposed. Only radial p modes were considered.

 \subsubsection{Stochastic excitation model}
 \label{sec:mod} 

The amplitudes of stochastically excited oscillations are obtained in
the manner of Chaplin et al. (2005). The procedure is based on the
formulation by Balmforth (1992b) and Goldreich \& Keeley (1977) but
includes a consistent treatment of the anisotropy parameter $\Phi$ of
the turbulent velocity field (for details see the discussion in
Chaplin et al. 2005). For the anisotropy parameter we adopt the value
$\Phi=1.13$, a value that was also considered by Chaplin et al. The
excitation model assumes a description in which the largest,
energy-bearing eddies are described by the mixing-length approach. The
small-scale convection is modelled by a turbulence spectrum, for which
we adopt the Kolmogorov spectrum (Kolmogorov 1941) describing the
spatial properties of the small-scale turbulence. The temporal
behaviour of the small-scale turbulent dynamics has a frequency
spectrum that is approximated by a Gaussian centred at zero frequency
and with a width corresponding to the inverse of the correlation
timescale of an eddy whose spatial extent (or wavenumber) is
characterized by the mixing length. The correlation timescale is not a
well-defined quantity, and therefore we scale a (well-defined)
characteristic eddy turnover time with a correlation parameter
$\lambda$ (cf. Balmforth 1992b), which we set to $\lambda=1$.

 \subsection{Mode peak parameters}
 \label{sec:params}

In this section we describe how the results of the stellar model
computations were used to make predictions of the parameters of the
p-mode peaks observed in the frequency power spectrum. For given
values of $a^2$ and $b^2$ the model computations provided predictions
of the linear damping rates, $\eta$, and the acoustic energy supply
rates, $P$, of the radial p modes. The observed parameters of the mode
peaks are formed from these quantities. The peak \textsc{fwhm}
linewidths are given by
 \begin{equation}
 \Delta = \eta / \pi.
 \label{eq:wid}
 \end{equation}
The mode velocity powers, $V^2$, are calculated from (e.g., Houdek et
al. 1999):
 \begin{equation}
 V^2 = \frac{P}{2\eta I} = \frac{P}{2\pi I \Delta},
 \label{eq:V2}
 \end{equation}
where $I$ is the mode inertia. The maximum power spectral density, or
height $H$, of a mode peak in the frequency power spectrum depends on
the effective length of the dataset, $T$, and the linewidth, $\Delta$,
via the formula (Fletcher et al. 2006):
 \begin{equation}
 H = \left(\frac{PT}{\eta I}\right) \frac{1}{\eta T + 2}
   = \frac{2V^2T}{\pi T \Delta +2}.
 \label{eq:height}
 \end{equation}
When $\pi T \Delta << 2$ (i.e., when $T << 2/\eta$) the mode is not
resolved and power is confined largely in one bin of the frequency
power spectrum, so that $H \sim V^2$. When, on the other hand, $\pi T
\Delta >> 2$ (i.e., $T >> 2/\eta$), the mode is well resolved and the
Lorentzian shape of the peak may be inferred from the power in the
bins occupied by the peak. For cases in the intermediate regime --
where $\pi T \Delta$ is neither much greater, or much smaller, than
two -- there is a gradual transition between $H$ for the
unresolved and fully resolved regimes.

In other recent papers (e.g., Chaplin et al. 2005) we have gone on to
use $H$ to make comparisons of theory with observation.  In this paper
we instead use data on heavily smoothed frequency power spectra, which
give direct inference on $V^2$ (and therefore the amplitudes, $V$) as
opposed to the $H$. However, we do still make use of the heights to
calibrate the model computations, as we now go on to discuss.

 \subsection{Notes on calibration}
 \label{sec:cal}

We make three important points concerning the calibration. Firstly,
the computations have been calibrated so that, for a model of the Sun,
the average maximum power spectral density, $H$, of the five most
prominent modes is the same as that observed in the real BiSON
Sun-as-a-star data. By using an average over several modes, as opposed
to taking the $H$ of just the strongest mode, we seek to stabilize the
calibration against small-scale fluctuations in the computations
(Chaplin et al. 2007). Furthermore, by using data on $H$, as opposed
to $V^2$ (or $V$), we maintain consistency with our previous work
(e.g., see Chaplin et al. 2008). To summarize our first point: data on
the Sun serve as a reference calibration for the model computations.

Secondly, it is important to recognise that the reference calibration
had to be performed independently for each of the non-local parameter
choices $a^2=b^2=600$ and $a^2=b^2=300$, respectively. This is because
the changes to the values of the non-local parameters can have a
significant impact on the computed damping rates, which in turn
affects the absolute magnitudes of the predicted velocity powers,
$V^2$ (Equation~\ref{eq:V2}) and therefore the heights, $H$
(Equation~\ref{eq:height}).

Thirdly, when it comes to comparing the results of the pulsation
computations with the observational data on $\alpha$\,Cen B, we must
remember that there will be instrument-dependent differences in the
observed velocity amplitudes, due to the use of different spectral
lines in the Doppler velocity observations. We have folded into the
calibration of the model data the fact that the amplitudes of the
solar p modes measured by the stellar techniques are a factor
1.07-times smaller than the amplitudes measured by BiSON (Kjeldsen et
al., 2008).

 \section{Results}
 \label{sec:res}

 \subsection{Data from the observations}
 \label{sec:obs}

\begin{figure*}

 \epsscale{1.0}
 \plottwo{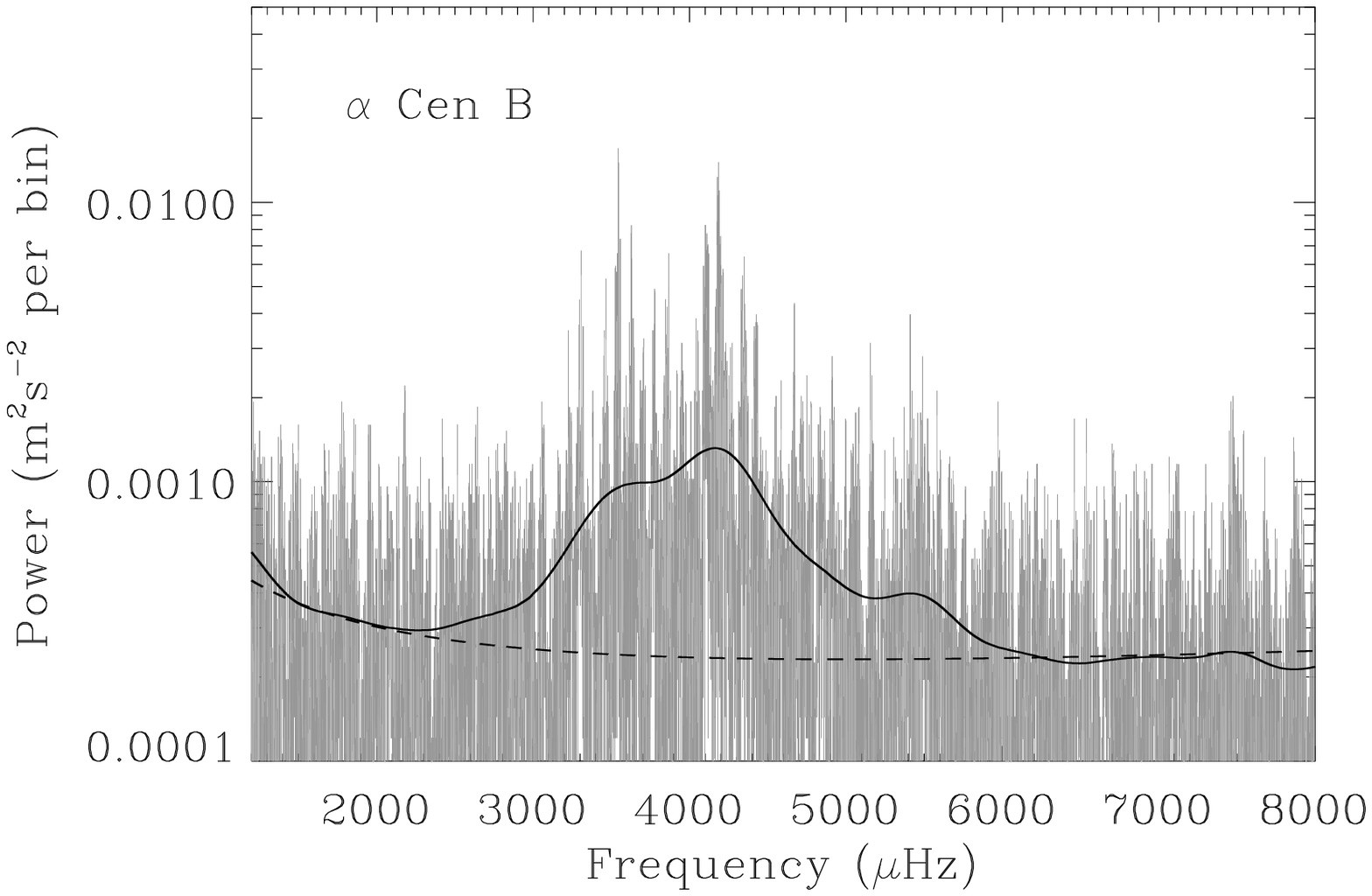}{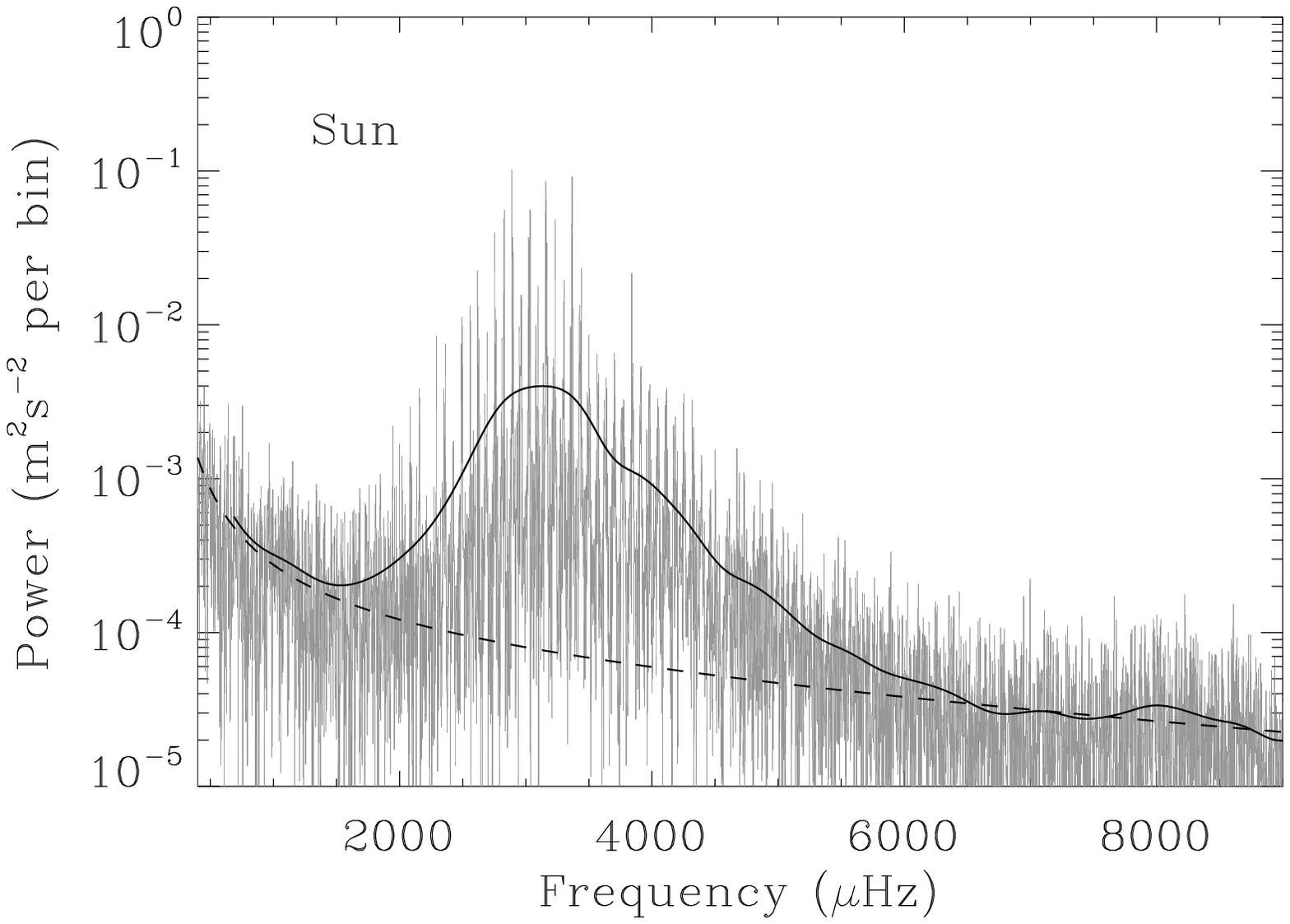}

 \caption{Left-hand panel: Observed frequency power spectrum of
 $\alpha$\,Cen~B (Kjeldsen et al. 2005). The dark solid line is the
 smoothed spectrum, while the dashed line is an estimate of the
 background. Right-hand panel: Observed frequency power spectrum of a
 5-day segment of BiSON Sun-as-a-star data (linestyles as per
 left-hand panel).}

 \label{fig:rawspecs}
 \end{figure*}

The left-hand panel of Fig.~\ref{fig:rawspecs} shows the observed
frequency power spectrum of $\alpha$\,Cen~B, plotted on a logarithmic
scale. The spectrum was computed from a few nights of Doppler velocity
data collected by Kjeldsen et al. (2005). The dark solid line is a
smoothed spectrum given by applying to the raw spectrum a Gaussian
filter of width $4\Delta\nu$, where $\Delta\nu$ is the large frequency
spacing between consecutive overtones (here $162\,\rm \mu Hz$).  The
dashed line is a smooth estimate of the background power spectral
density. It was obtained by fitting, in regions outside the range
occupied by the p modes, a second-order polynomial to the logarithm of
power versus the logarithm of frequency.

We used the Gaussian smoothed spectrum and the background fit to
estimate the mode amplitudes, $V$. This was done by following the
recipe outlined in Kjeldsen et al. (2008)\footnote{Note that we get
mode amplitudes for $\alpha$\,Cen~B and the Sun that are a few percent
lower than in Kjeldsen et al. These differences arise from differences
in the fitting-function that was used to estimate the background. The
estimation of the background is the largest source of uncertainty for
the method.}. In summary, we began by subtracting the background fit
from the Gaussian-smoothed spectrum. The residuals thereby obtained
were converted to units of power per Hertz, multiplied by the large
frequency spacing $\Delta\nu$, and finally divided by a constant
factor (see factors in Table~1 of Kjeldsen et al. 2008) to allow for
the effective number of modes in each slice $\Delta\nu$ of the
spectrum.  This gave observational estimates of the radial mode
amplitudes, $\rm V_{obs}$, which are plotted as a thick grey line in
each panel of Fig.~\ref{fig:cenamp}. The thin grey lines are an
estimate of the uncertainty envelope on the observed amplitudes. These
uncertainties were estimated by using as a guide the results of
analyzing many independent, short segments of BiSON Sun-as-a-star
data, as will be explained below.

We have used results on a parallel analysis of the BiSON Sun-as-a-star
data as a belt-and-braces check on the procedures and results.  The
complete BiSON timeseries that we used is 4752\,days long. This was
split into independent segments of length 5\,days (a length similar to
the $\alpha$\,Cen~B timeseries), and the analysis procedures that were
applied to the $\alpha$\,Cen~B frequency power spectrum were applied
to the frequency power spectrum of each segment. The right-hand panel
of Fig.~\ref{fig:rawspecs} shows the frequency power spectrum of one
of the 5-day BiSON segments (the layout and linestyles are the same as
in the left-hand panel, and the spectrum was again smoothed over
$4\Delta\nu$, but with $\Delta\nu=135\,\rm \mu Hz$ for the Sun).

The thick grey lines in both panels of Fig.~\ref{fig:sunamp} show the
\emph{mean} of the amplitudes that were obtained from the $\sim 950$
independent 5-day BiSON segments, using the method of Kjeldsen et
al. outlined above.  The thin grey lines mark plus and minus the rms
of the amplitudes. Since the total epoch covered by the data spans
more than one 11-yr cycle of solar activity, these uncertainties
reflect variation of the estimated amplitudes from short-term
stochastic variability \emph{and} long-term solar cycle
variability. There will also be a contribution from the finite
signal-to-noise ratio of the observations.

The fractional uncertainties in the estimated 5-day BiSON amplitudes
are about 15\,\% in the middle of the spectrum; and about 25\,\% at
the extreme frequencies $\sim 2000$ and $\sim 5000\,\rm \mu Hz$,
respectively. In the light of these values, we have assumed the
estimated $\alpha$\,Cen B amplitudes are all determined to a
fractional precision of 20\,\%. This value was used to make the
uncertainty envelope for the estimated $\alpha$\,Cen B amplitudes
plotted in both panels of Fig.~\ref{fig:cenamp}.

With estimates of the uncertainties on $\rm V_{obs}$ now in hand, next
we ask the question: how robust are the $\rm V_{obs}$ likely to be for
$\alpha$\,Cen~B?  We can obtain some insight by testing the robustness
of the 5-day BiSON Sun-as-a-star amplitudes. This test may be
accomplished by comparing the 5-day estimates with estimates from a
``peak-bagging'' analysis of the entire 4752-day BiSON timeseries. The
BiSON mode amplitudes are usually estimated using the peak-bagging
fitting techniques (e.g., see Chaplin et al. 2006). Peak-bagging
involves maximum-likelihood fitting of mode peaks in the frequency
power spectrum to multi-parameter fitting models, where individual
mode peaks are represented by Lorentzian-like functions. The points
with error bars in Fig.~\ref{fig:sunamp} show estimated amplitudes
$\rm V_{bag}$ from a full peak-bagging analysis of the 4752-day BiSON
timeseries, in which the mode peaks were fitted in the high-resolution
frequency power spectrum of the complete timeseries. Since these
peak-bagging estimates are extremely precise, and are also expected to
be fairly accurate, they serve to provide a robust cross-check of the
5-day estimates. As we can see, the 5-day BiSON estimates are in good
agreement with the peak-bagging BiSON estimates. This comparison
suggests we may also expect to have reasonable confidence in the
observed $\alpha$\,Cen B amplitudes.

 \subsection{Comparison of theory with observation}
 \label{sec:test}

\begin{figure*}
 \epsscale{1.0}
 \plottwo{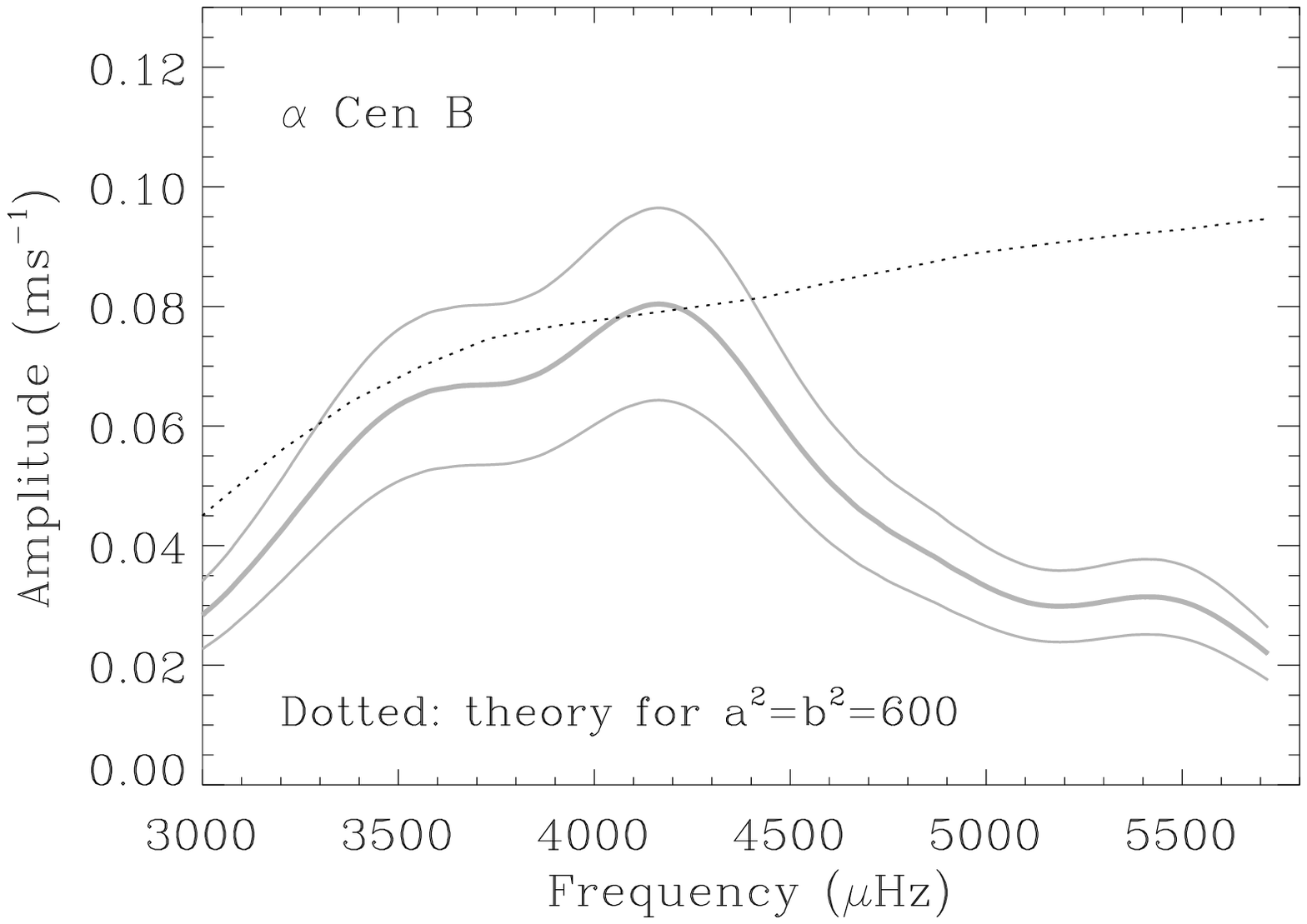}{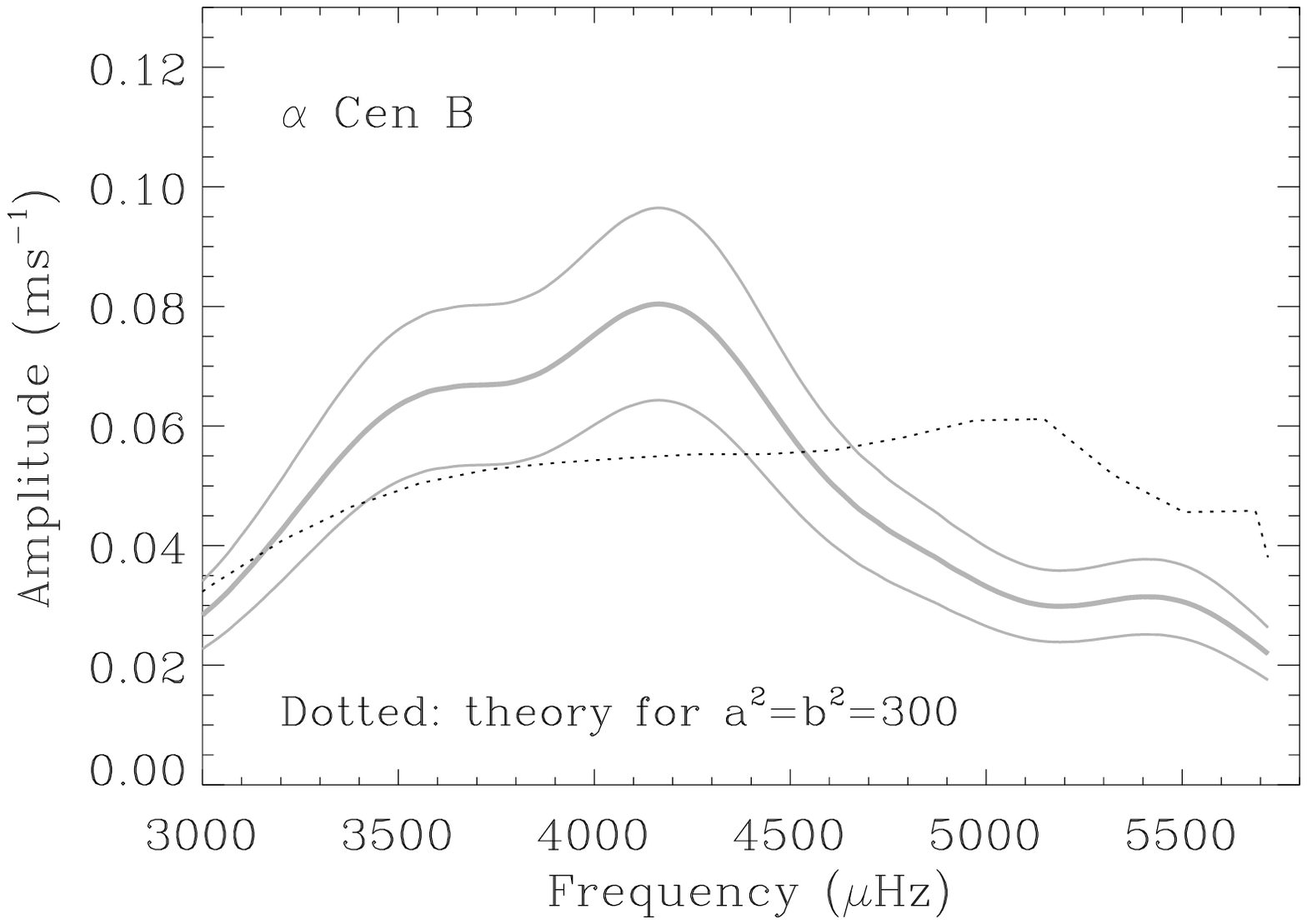}

 \caption{Observed mode velocity amplitudes $\rm V_{obs}$ (thick grey
 line) and theoretically computed amplitudes $V$ (dotted line) of
 $\alpha$\,Cen B. The thin grey lines denote the estimated uncertainty
 envelope on the amplitudes (see text). Left-hand panel: theoretical
 predictions for $a^2=b^2=600$. Right-hand panel: theoretical
 predictions for $a^2=b^2=300$.}

 \label{fig:cenamp}
 \end{figure*}

\begin{figure*}
 \epsscale{1.0}
 \plottwo{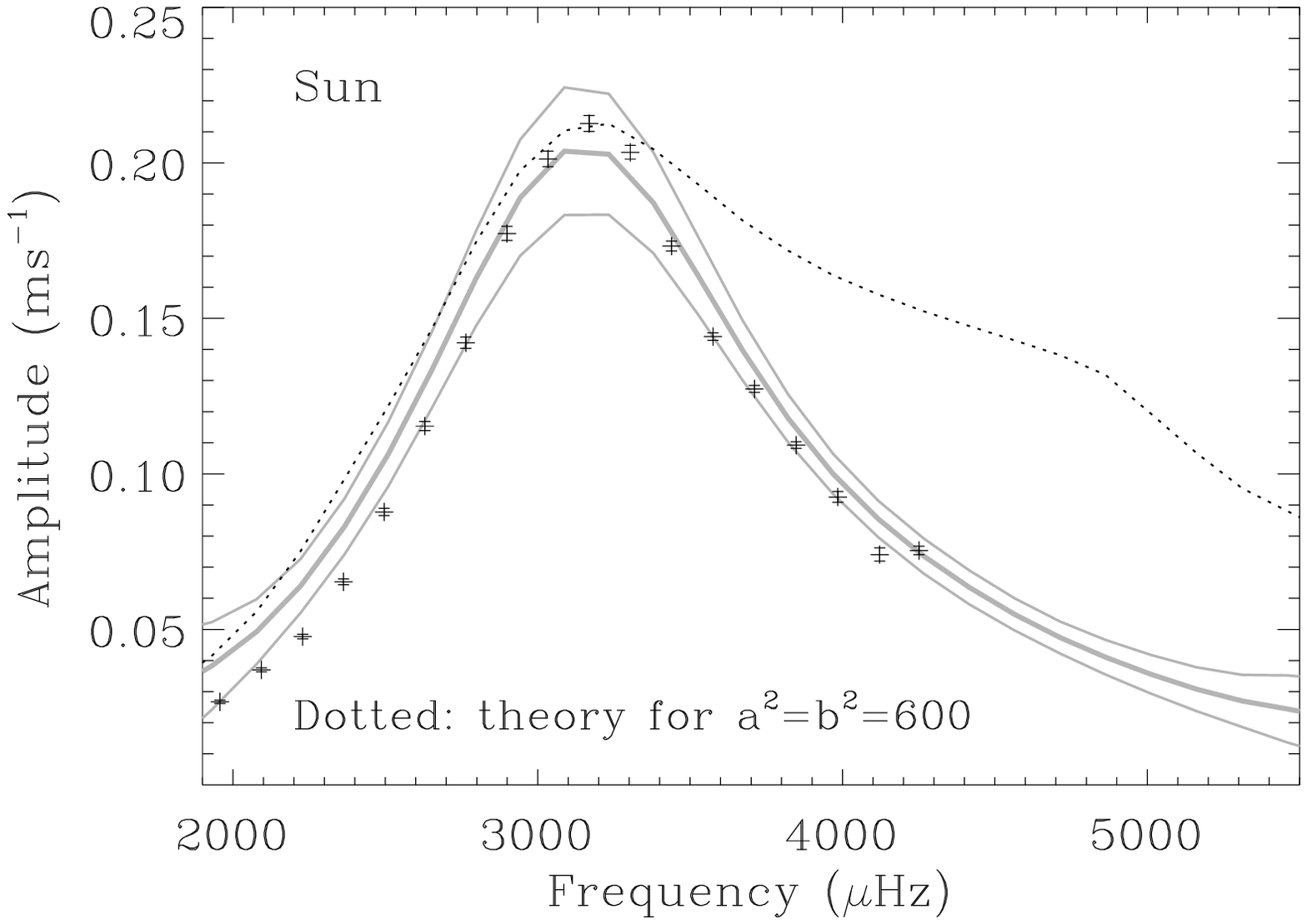}{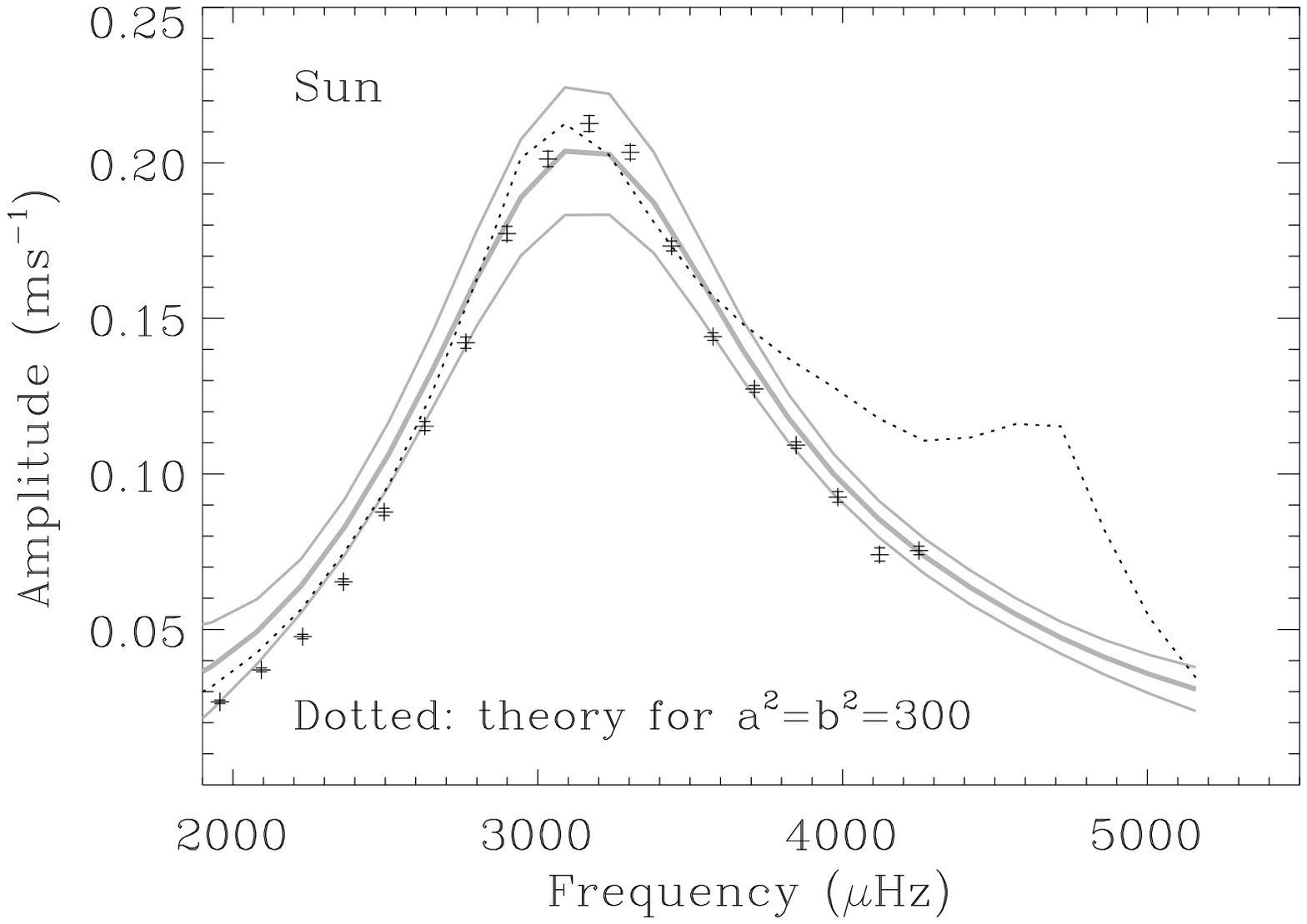}

 \caption{Observed mode velocity amplitudes $\rm V_{obs}$ (thick grey
 line) and theoretically computed amplitudes $V$ (dotted line) of the
 Sun. The plotted amplitudes are the mean amplitudes from analyzing
 950 independent 5-day segments of BiSON Sun-as-a-star data. The thin
 grey lines denote the uncertainty envelope on the amplitudes, from
 the scatter in the results from the 5-day segments.  The symbols with
 error bars are amplitudes $\rm V_{bag}$ given by a ``peak-bagging''
 analysis of the full 4752-day BiSON timeseries. Left-hand panel:
 theoretical predictions for $a^2=b^2=600$. Right-hand panel:
 theoretical predictions for $a^2=b^2=300$.}

 \label{fig:sunamp}
 \end{figure*}

Predictions from the pulsation computations of the p-mode velocity
amplitudes, $V$, are shown as dotted lines in Fig.~\ref{fig:cenamp}
($\alpha$\,Cen B) and Fig.~\ref{fig:sunamp} (Sun). The predicted
amplitudes plotted in the left-hand panels of the figures are for
$a^2=b^2=600$, while those in the right-hand panels are for
$a^2=b^2=300$.

First, let us compare the modelled and observed amplitudes for
$\alpha$\,Cen B (Fig.~\ref{fig:cenamp}). When $a^2=b^2=600$ (left-hand
panel) the model amplitudes form a very broad plateau at frequencies
$\ga 3700\,\rm \mu Hz$, and the level of this plateau rises slowly
with increasing frequency. While the match between the predicted
amplitudes and the observed amplitudes is reasonable at frequencies
$\la 4500\,\rm \mu Hz$, this is demonstrably not so at higher
frequencies, where the predicted amplitudes are significantly higher
than the observed amplitudes. When $a^2=b^2=300$ the predicted
amplitudes give a better match, on average, to the observed
amplitudes, but some overestimation at higher frequencies remains.

High-frequency disagreement between theory and observation is also
seen in Fig.~\ref{fig:sunamp} for the Sun. When $a^2=b^2=600$, we
again see a pronounced disagreement at high frequencies between the
predicted amplitudes and the observed amplitudes. When $a^2=b^2=300$
the level of disagreement is less pronounced, but is nevertheless
still present.

Let us summarize the main points from Figs.~\ref{fig:cenamp}
and~\ref{fig:sunamp}: when $a^2=b^2=600$, theoretically computed
high-frequency velocity amplitudes for $\alpha$\,Cen B \emph{and} the
Sun significantly overestimate the observed velocity
amplitudes. Although not as severe, this overestimation persists at
$a^2=b^2=300$. What might be the cause of this disagreement between
theory and observation? That is the question we turn to next.

 \section{Inference on the p-mode linewidths}
 \label{sec:wid}

Provided we trust the mode amplitudes $\rm V_{obs}$ estimated from the
$\alpha$\,Cen~B frequency power spectrum -- and the 5-day BiSON
analysis above suggests the estimates should be reasonable --
there are two possible ways out of the problem. Equation~\ref{eq:V2}
implies that:
 \begin{equation}
 \frac{\delta V}{V} \sim \frac{\delta P}{2P} - \frac{\delta\eta}{2\eta}.
 \label{eq:changes}
 \end{equation}
So, either the model damping rates $\eta$ -- and therefore the model
linewidths $\Delta$ -- are too weak at high frequencies to explain the
observed $\rm V_{obs}$, or the modelled acoustic supply rates, $P$, are too
strong (or there is a combination of the two effects).

There is good evidence from analysis of Sun-as-a-star data (e.g.,
Chaplin et al. 2005; Houdek 2006) that a significant part of the
disagreement may come from problems computing $\eta$ (and therefore
$\Delta$).  We can double-check the validity of this assumption here
for the Sun, because we have precise BiSON data on the linewidths and
powers of the solar p modes from the peak-bagging analysis. These
peak-bagging data may be used as a precise and accurate reference
against which to check the quality of the results obtained on the
short 5-day BiSON segments.

 \subsection{The Sun-as-a-star linewidths}
 \label{sec:sas}

\begin{figure*}
 \epsscale{1.0}
 \plottwo{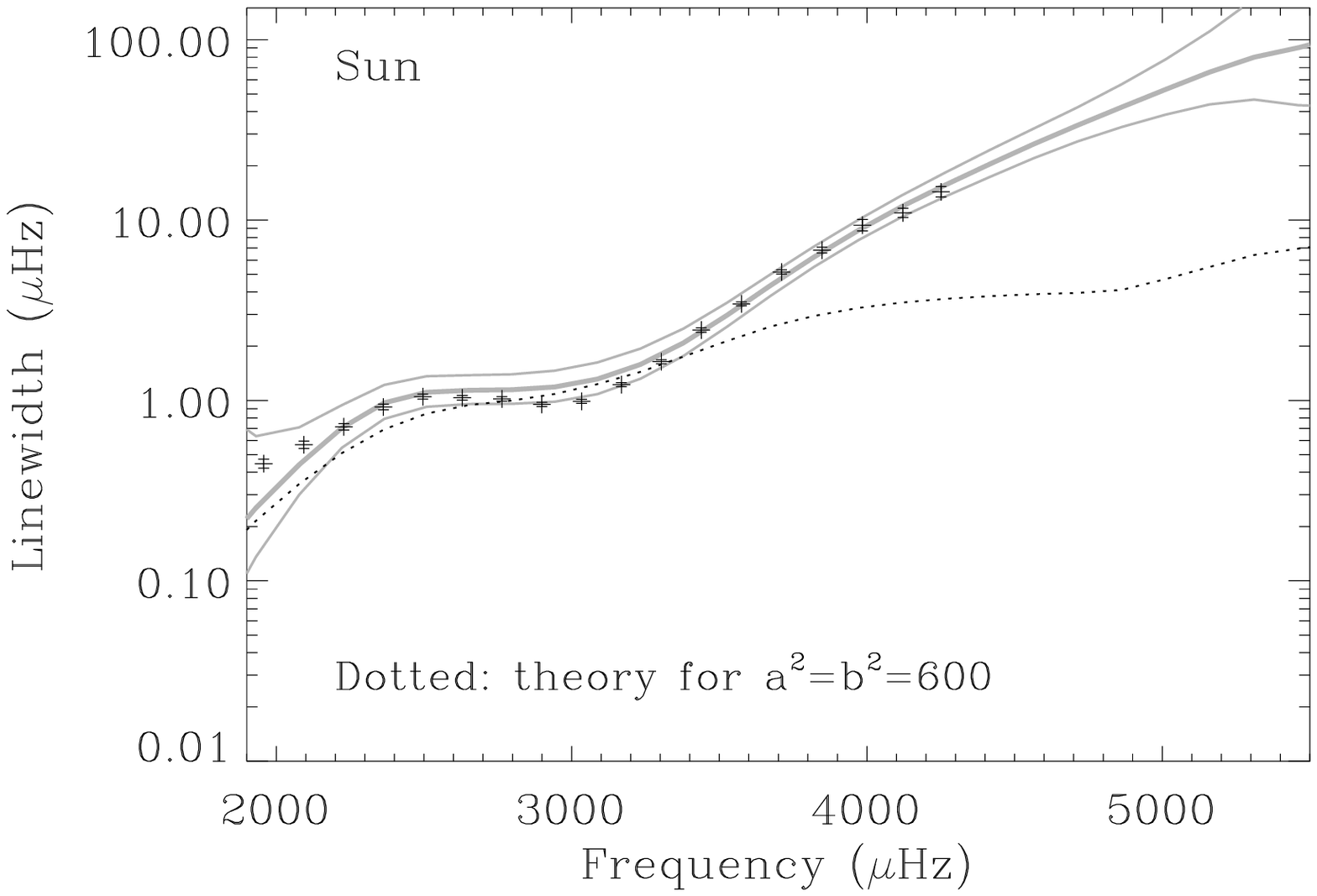}{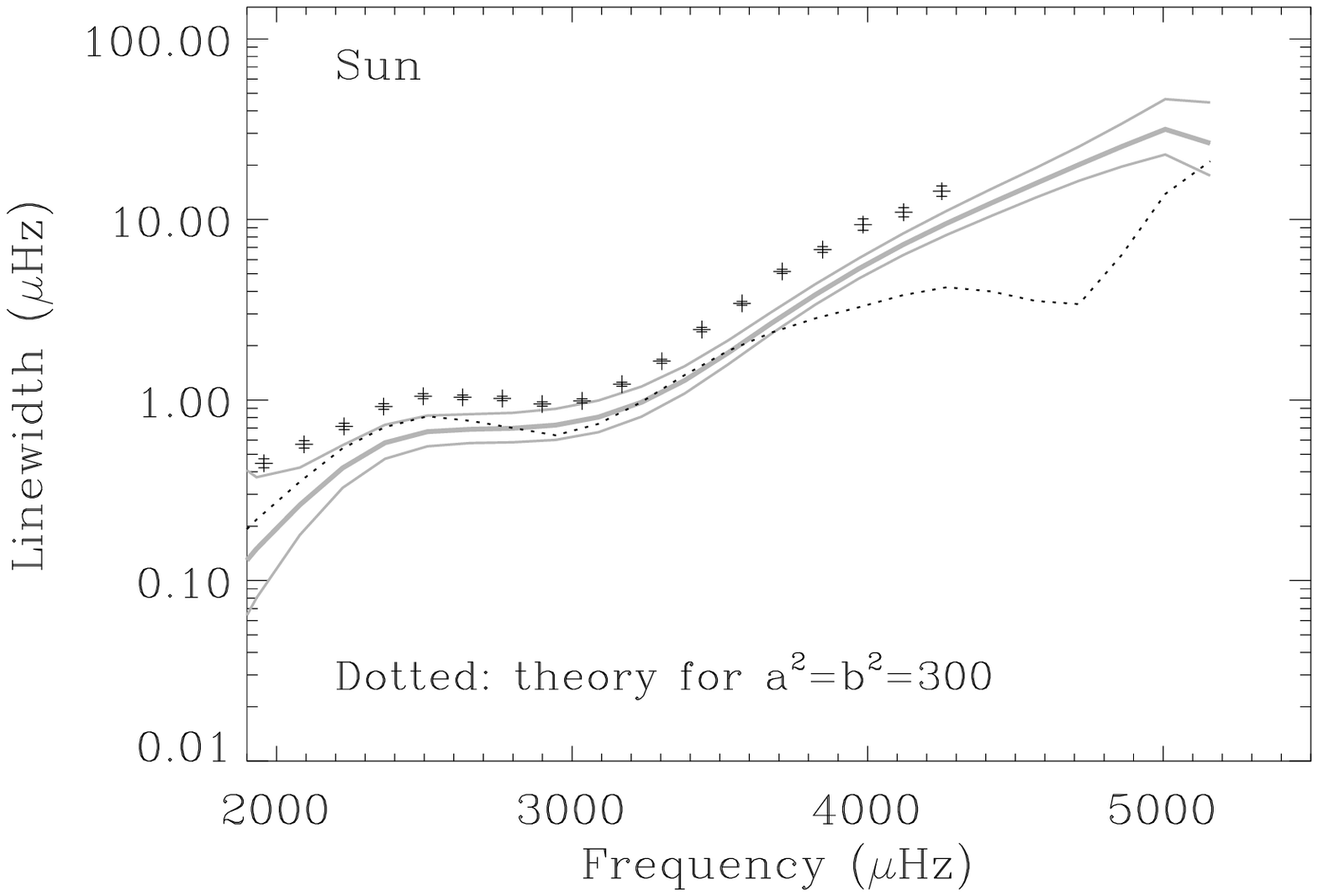}

 \caption{Inferred linewidths $\rm \Delta_{obs}$ (thick grey line) and
 theoretically computed linewidths $\Delta$ (dotted line) of the
 Sun. The theoretically computed linewidths have been smoothed with a
 median smoothing filter of three consecutive values, as in Houdek \&
 Gough (2002). The plotted inferred linewidths are the mean inferred
 linewidths from analyzing 950 independent 5-day segments of BiSON
 Sun-as-a-star data. The thin grey lines denote the uncertainty
 envelope on the inferred linewidths, from the scatter in the results
 from the 5-day segments.  The symbols with error bars are linewidths
 $\rm \Delta_{bag}$ given by a ``peak-bagging'' analysis of the full
 4752-day BiSON timeseries. Left-hand panel: theoretical predictions
 for $a^2=b^2=600$. Right-hand panel: theoretical predictions for
 $a^2=b^2=300$.}

 \label{fig:sunwid}
 \end{figure*}

We base our solar check on Equation~\ref{eq:V2}. Again, it tells us
that $V^2 \propto P / \Delta$. If we take the ratio of the observed to
the model-computed velocity powers, we will therefore have:
 \begin{equation}
 \left(\frac{\rm V^2_{obs}}{V^2}\right) = 
 \left(\frac{\rm P_{obs}}{P}\right)
 \left(\frac{\Delta}{\rm \Delta_{obs}}\right).
 \label{eq:ratio}
 \end{equation}
In the above, all observed quantities for the star, which we assume
come from analysis of the short 5-day segments, carry the suffix
``obs''; and the model-predicted quantities are
suffix-free. Rearrangement of the above gives:
 \begin{equation}
 {\rm \Delta_{obs}} =   \left(\frac{\rm P_{obs}}{P}\right)
                        \left(\frac{V^2}{\rm V^2_{obs}}\right) 
			\Delta.
 \label{eq:ratio1}
 \end{equation}
Let us suppose that all the problems in the modelling lie in
computation of the linewidths, $\Delta$. The implication is then, of
course, that the model-predicted values of the acoustic supply rates,
$P$, are accurate, i.e., $P=\rm P_{obs}$. Equation~\ref{eq:ratio1}
would then simplify to
 \begin{equation}
 {\rm \Delta_{obs}} =   \left(\frac{V^2}{\rm V^2_{obs}}\right) 
			\Delta.
 \label{eq:ratio2}
 \end{equation}
How well do the solar model computations and 5-day observations match
Equation~\ref{eq:ratio2}, i.e., how well is the relation $P=\rm
P_{obs}$ satisfied? Fig.~\ref{fig:sunwid} shows plots of the inferred
linewidths $\rm \Delta_{obs}$ of the Sun (thick grey line). The $\rm
\Delta_{obs}$ were computed from the observed 5-day $\rm V_{obs}$ in
Section~\ref{sec:obs} above and the model predictions of $V^2$ and
$\Delta$. The left-hand panel shows results for when the model
computations use $a^2=b^2=600$; while the right-hand panel shows
results with $a^2=b^2=300$. The thin grey lines show the uncertainty
envelopes on the $\rm \Delta_{obs}$, which come from the uncertainties
on $\rm V_{obs}$ shown in Fig.~\ref{fig:sunamp}.

Also plotted as points with associated best-fitting uncertainties in
Fig.~\ref{fig:sunwid} are the fitted linewidths $\rm \Delta_{bag}$
from a full 4752-day peak-bagging analysis of the BiSON Sun-as-a-star
data. For our purposes here, we may regard the peak-bagging linewidths
as being good measures of the true linewidths of the Sun. Our check on
whether $P=\rm P_{obs}$ therefore amounts to seeing if the 5-day $\rm
\Delta_{obs}$ are a good match to the $\rm \Delta_{bag}$. This does
assume that there are no significant biases in the estimated $\rm
V_{obs}$, as was suggested by the good agreement of the $\rm V_{obs}$
and $\rm V_{bag}$ in Fig.~\ref{fig:sunamp}.

\begin{figure*}
 \epsscale{1.0}
 \plotone{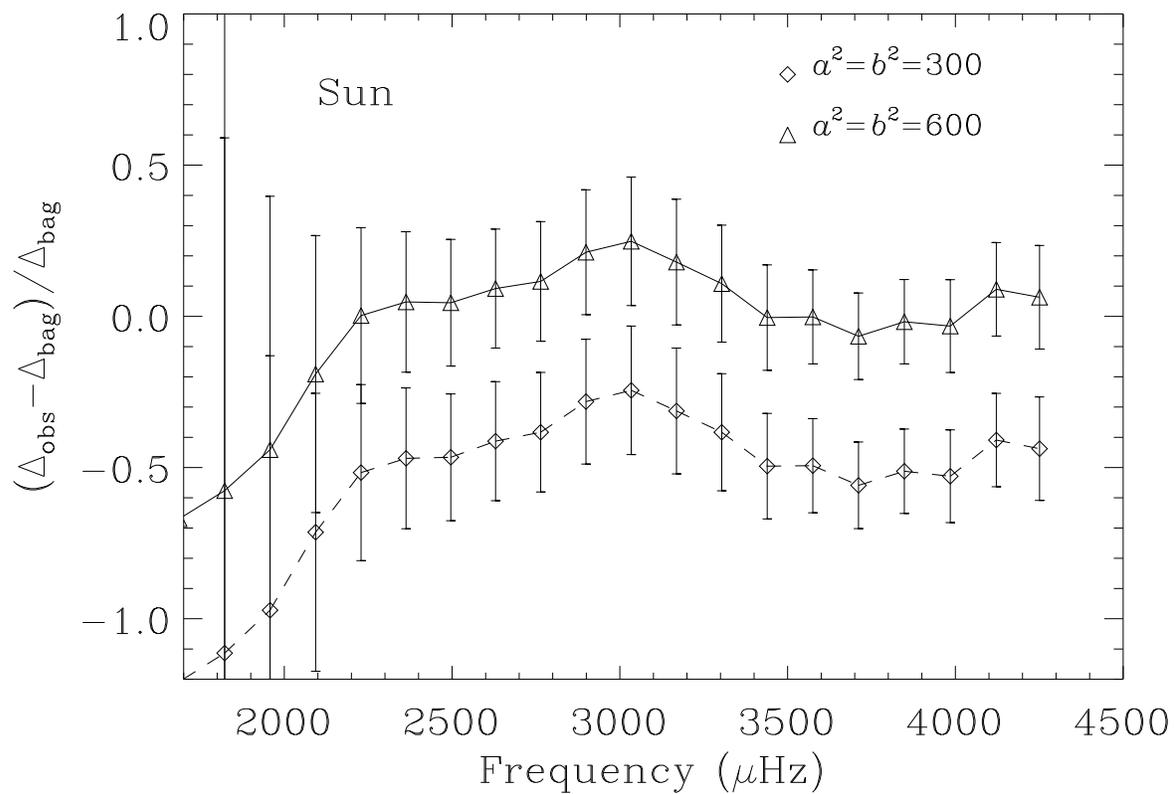}

 \caption{Results on the solar p-mode linewidths. Plotted are
  differences in the natural logarithms (i.e., absolute fractional
  differences) of the inferred linewidths, $\rm \Delta_{obs}$ (which
  use data from the 5-day BiSON segments) and peak-bagging linewidths,
  $\rm \Delta_{bag}$ (from the frequency power spectrum of the full
  4752-day BiSON timeseries).}

 \label{fig:widrat}
 \end{figure*}

At a first glance, the most striking aspect of the solar plots is
indeed the encouraging level of agreement between the $\rm
\Delta_{obs}$ and the $\rm \Delta_{bag}$. Fig.~\ref{fig:widrat} shows
the comparison in more detail. Here, we have plotted the fractional
differences between the 5-day linewidths and the peak-bagging
linewidths. When $a^2=b^2=600$, the $\rm \Delta_{obs}$ over the main
part of the solar p-mode spectrum are seen to be about 10\,\% higher
on average than the $\rm \Delta_{bag}$, while they are about 35\,\%
lower when $a^2=b^2=300$. The implication is that in both cases there
is actually an offset between $P$ and $\rm P_{obs}$. However, what we
can say is that changes in the differences as a function of frequency
are not significant over the main part of the p-mode spectrum, given
the observational uncertainties (we can disregard the differences at
the lowest frequencies, where the error bars are very large, and the
5-day spectra have insufficient resolution to give robust estimates of
$\rm \Delta_{obs}$ in this part of the spectrum). We are therefore in
a position to conclude the following: the \emph{shape} in frequency of
the $\rm \Delta_{obs}$ is a reasonable match to the shape of the $\rm
\Delta_{bag}$ at the level of precision of 5-day timeseries.

The above suggests we may use Equation~\ref{eq:ratio2} to infer the
variation of the linewidths as a function of frequency, without the
need to measure those linewidths directly.  There will remain some
uncertainty over the absolute calibration of the linewidths. Let us
now apply Equation~\ref{eq:ratio2} to the data on $\alpha$\,Cen~B.

 \subsection{Inference on linewidths of $\alpha$\,Cen~B}
 \label{sec:acenb}

\begin{figure*}
 \epsscale{1.0}
\plottwo{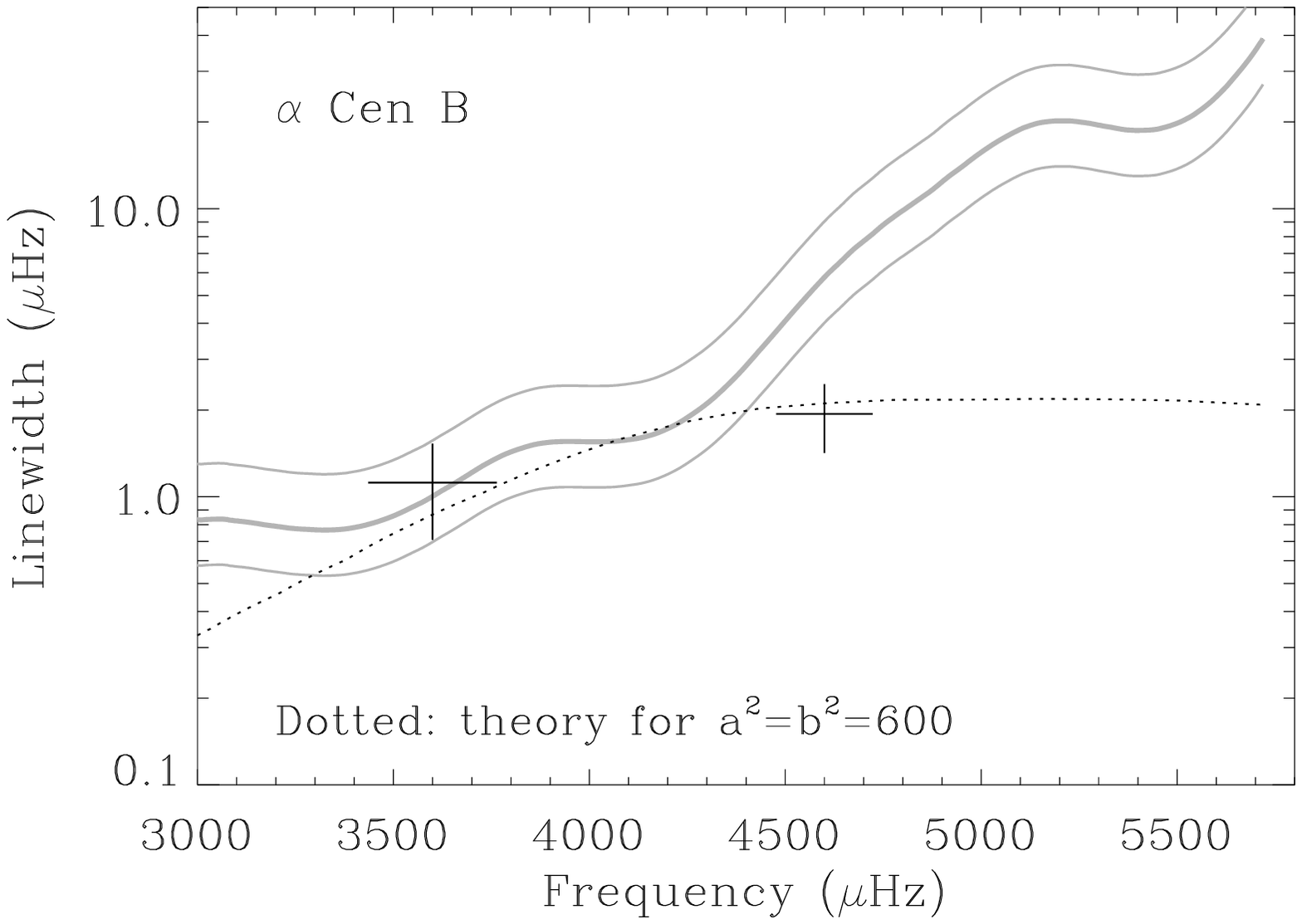}{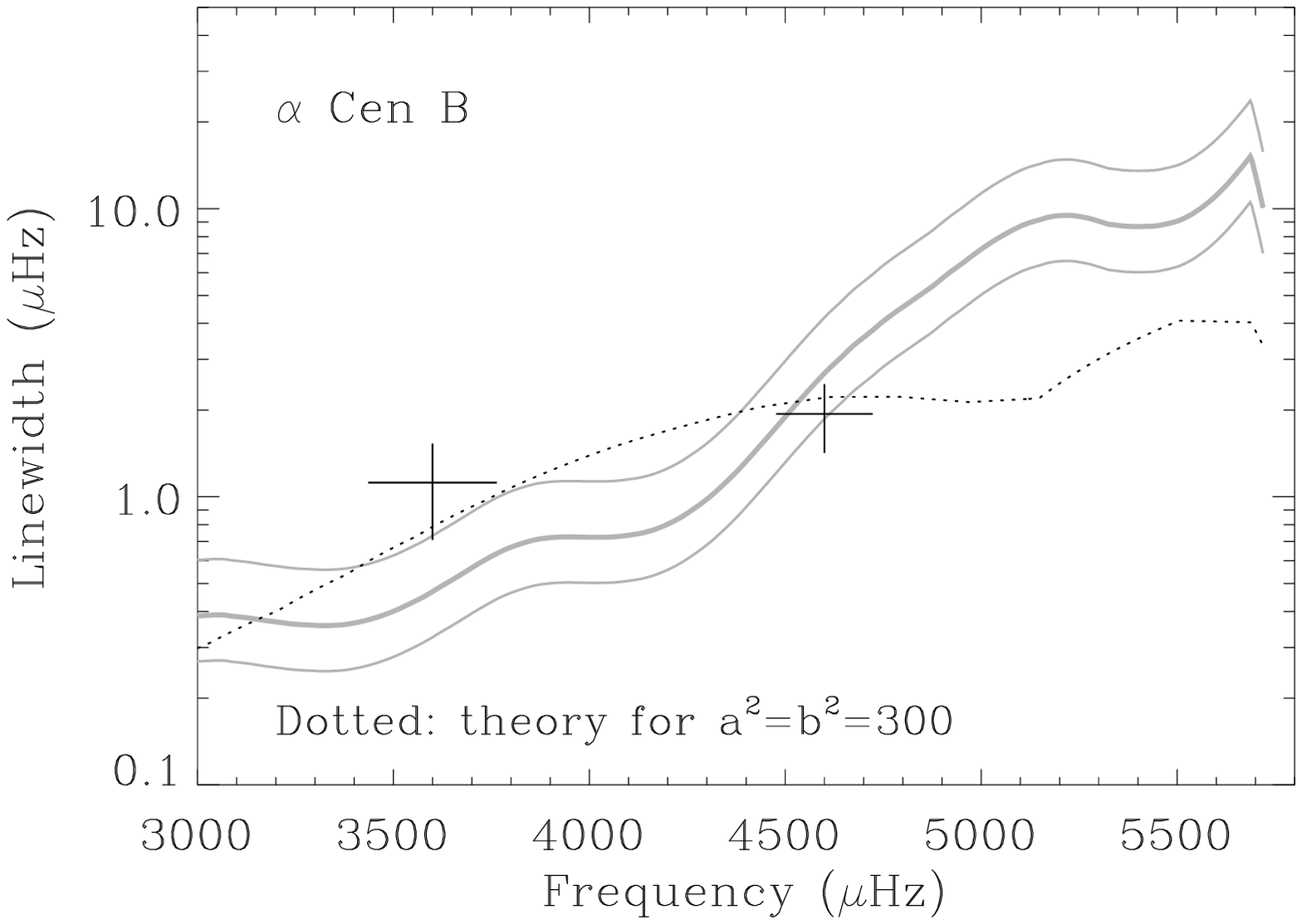}

 \caption{Inferred linewidths $\rm \Delta_{obs}$ (thick grey line) and
 theoretically computed linewidths $\Delta$ (dotted line) of
 $\alpha$\,Cen B. The theoretically computed linewidths have been
 smoothed with a median smoothing filter of three consecutive values,
 as in Houdek \& Gough (2002). The thin grey lines denote the
 estimated uncertainty envelope on the inferred linewidths (see
 text). The symbols with error bars show observational linewidth
 estimates from Kjeldsen et al. (2005). Left-hand panel: theoretical
 predictions for $a^2=b^2=600$. Right-hand panel: theoretical
 predictions for $a^2=b^2=300$.}

 \label{fig:cenwid}
 \end{figure*}

We have used the observed estimates of $\rm V_{obs}$ from the
$\alpha$\,Cen B spectrum, together with the theoretical computations
of $V$ and $\Delta$ for the star, to give the inferred linewidths $\rm
\Delta_{obs}$ plotted in Fig.~\ref{fig:cenwid}. We do not require an
explicit estimate of $\rm P_{obs}$, because we rely on the assumption
(verified above for the Sun) that the theoretical $P$ has the same
shape in frequency as $\rm P_{obs}$.

There is no obvious reason why we should expect the acoustic
supply-rate computations, $P$, to be any less valid for $\alpha$\,Cen
B than they are for the Sun (this might not have been so had
$\alpha$\,Cen B been somewhat hotter than the Sun, or much cooler than
it actually is; see Houdek 2006). At the very least we may therefore
have reasonable confidence in the shapes of the inferred linewidth
curves for the star, assuming any deviations from a constant scaling
offset are no more severe than they are for the Sun. Both curves in
Fig.~\ref{fig:cenwid} appear to show a plateau in the linewidths at
$4000\,\rm \mu Hz$, like that seen in the solar linewidth curve. Other
obvious features of the curves -- a decrease of the linewidths at
lower frequencies, and an increase at higher frequencies -- are also
Sun-like in nature. It is worth stressing that the linewidth curves
have very similar shapes at $a^2=b^2=600$ (left-hand panel) and
$a^2=b^2=300$ (right-hand panel).

We have also shown on Fig.~\ref{fig:cenwid} observational linewidth
estimates, which were obtained by Kjeldsen et al. (2005) in two parts
of the p-mode spectrum (points with associated uncertainties). These
estimates were obtained from the same data but in a different way,
namely by measuring the scatter of p-mode frequencies around smooth
ridges in the echelle diagram. Comparison of the inferred $\rm
\Delta_{obs}$ with the Kjeldsen et al. estimates allows us to place
constraints on the absolute values of the linewidths for
$\alpha$\,Cen~B. Such a comparison suggests there is actually little
to choose between the $\rm \Delta_{obs}$ and the Kjeldsen et
al. estimates when the energy supply rates are computed using
$a^2=b^2=600$ (giving the inferred linewidths in the left-hand panel
of Fig.~\ref{fig:cenwid}) or $a^2=b^2=300$ (giving the inferred
linewidths in the right-hand panel). However, the $a^2=b^2=300$ result
does give a slightly better match. Inspection of the $\rm
\Delta_{obs}$ linewidth curve then implies that the $\alpha$\,Cen~B
linewidths take values of about $0.4\,\rm \mu Hz$ at a frequency of
$3500\,\rm \mu Hz$, about $0.8\,\rm \mu Hz$ at a frequency of
$4000\,\rm \mu Hz$, and about $2.5\,\rm \mu Hz$ at a frequency of
$4500\,\rm \mu Hz$. These linewidths turn out to agree reasonably well
with the solar linewidths, if the mode frequencies are multiplied by
the ratio of the acoustic cut-off frequencies of the two stars (which
in effect scales the $\alpha$\,Cen~B frequencies down to those shown
by the Sun). In short, the $\alpha$\,Cen~B linewidth curve is similar
to a frequency-scaled version of its solar cousin.

 \section{Discussion}
 \label{sec:disc}

In this paper we presented an analysis of the amplitudes and
linewidths of the low-degree, Sun-like p modes displayed by the star
$\alpha$\,Cen~B. These data were extracted from only a few nights of
Doppler velocity observations collected on the star by Kjeldsen et
al. (2005), and were compared with theoretical predictions of the
stochastic excitation and damping rates of the p modes. We also
performed a parallel analysis of Sun-as-a-star p-mode data collected
by the ground-based BiSON. The very long BiSON timeseries allowed us
to validate the analysis techniques.

For the Sun, we found that model predictions of the mode amplitudes
were significantly larger than the observed amplitudes in the
high-frequency part of the p-mode spectrum. We were able to
confirm that most of the disagreements for the Sun, which set in at
frequencies $\ga 3300\,\rm \mu Hz$, are due to problems computing the
damping rates, not the excitation rates, of the modes.  The computed
damping rates must be increased to explain the observed amplitudes.

Similar disagreements are seen for $\alpha$\,Cen~B. Here, we do not
have the luxury of cross-checking the analysis with results from very
long datasets, since the latter at present do not exist [although we
may hope to obtain such datasets in the future from the likes of SONG
(Grundahl et al. 2007) and SIAMOIS (Mosser et al. 2007)]. However, by
assuming that the disagreements for $\alpha$\,Cen~B are also due
largely to problems with the model-predicted damping rates, as was
shown to be the case for the Sun, we were able to demonstrate that the
model linewidths must also be increased significantly in order to
explain the observed amplitudes. The problems for $\alpha$\,Cen~B set
in at frequencies $\ga 4500\,\rm \mu Hz$.

The conclusions above bear on a prediction of the pulsation
computations mentioned in the Introduction (Section~\ref{sec:intro}):
that stars cooler than about 5400\,K will show a broad plateau or
double-hump of power in their p-mode spectra. We showed that
$\alpha$\,Cen~B does not have the predicted broad plateau in its
\emph{observed} p-mode spectrum. If the high-frequency damping rates
are increased -- as they must be to resolve the disagreement between
theory and observation -- the broad plateau or second high-frequency
hump disappears in the cooler models.

Finally, we also showed how, by making suitable comparisons of theory
and observation, it is possible to make inference on the variation with
frequency of the p-mode linewidths without directly measuring those
linewidths, even if the observations come from only a few days of
data. Use of independent measures of the $\alpha$\,Cen~B linewidths,
made in two parts of its spectrum by Kjeldsen et al. (2005), allowed
us to calibrate our inferred linewidth curve for $\alpha$\,Cen~B. We
found that the resulting, calibrated linewidth curve is similar to a
frequency-scaled version of its solar cousin, with the scaling factor
equal to the ratio of the respective acoustic cut-off frequencies of
the two stars. The ratio of the frequencies at which the onset of
high-frequency problems is seen in both stars is also given
approximately by the same scaling factor.

\acknowledgments

WJC thanks members of the School of Physics at the University of
Sydney for their hospitality and support during a visit when some of
this work was conducted. WJC also acknowledges the support of the
School of Physics and Astronomy at the University of Birmingham. GH
acknowledges the support of the UK Science and Technology Facilities
Council, and TRB acknowledges the support of the Australian Research
Council.


\end{document}